\documentclass[aps,prl,twocolumn,a4paper,10pt,notitlepage,footinbib,superscriptaddress,showpacs]{revtex4-1}%
\usepackage[english]{babel}
\usepackage[T1]{fontenc}
\usepackage{amssymb,amsmath,amsfonts}
\usepackage{textcomp}
\usepackage{graphicx,color}
\usepackage[dvipsnames]{xcolor}
\usepackage[utf8x]{inputenc}
\usepackage{bm}
\usepackage{float}
\usepackage{url}

\usepackage{tabularx}

\newcommand{\dmi}[1]{\textcolor{black}{#1}}%

\begin{document}

\title{Separation of Geometrical and Topological Entanglement in Confined Polymers Driven Out-Of-Equilibrium}

\author{Davide Michieletto}
\thanks{Joint first author}
\affiliation{School of Physics and Astronomy, University of Edinburgh, Peter Guthrie Tait Road, Edinburgh, EH9 3FD, United Kingdom}
\affiliation{MRC Human Genetics Unit, Institute of Genetics and Molecular Medicine, University of Edinburgh, North Crewe Rd, Edinburgh, EH4 2XU, United Kingdom}

\author{Enzo Orlandini$^\dagger$}
\thanks{Joint first author}
\affiliation{Dipartimento di Fisica e Astronomia and Sezione INFN, Universit\'a degli Studi di Padova, I-35131 Padova, Italy}

\author{Matthew S. Turner}
\affiliation{Department of Physics and Centre for Complexity Science, University of Warwick, Coventry, CV4 7AL, UK}
\affiliation{Department of Chemical Engineering, Kyoto University, Kyoto, Japan}

\author{Cristian Micheletti}
\affiliation{SISSA (Scuola Internazionale Superiore di Studi Avanzati), Via Bonomea 265, 34136 Trieste, Italy}

\begin{abstract}
\vspace{-0.3 cm}
We use Brownian dynamics simulations and advanced topological profiling methods to characterize the out-of-equilibrium evolution of self-entanglement in \dmi{linear} polymers confined into nano-channels and under periodic compression. \dmi{We distinguish two main forms of entanglement, geometrical and topological. The latter is measured by the number of (essential) crossings of the physical knot detected after a suitable bridging of the chain termini. The former is instead measured as the average number of times a linear chain appears to cross itself when viewed under all projections, and is irrespective of the physical knotted state.} The key discovery of our work is that these \dmi{two forms of entanglement} are uncoupled and evolve \dmi{with} distinct dynamics. While geometrical entanglement is typically in phase with the \dmi{compression-elongation} cycles and it is primarily sensitive to its force $f$, the topological measure is mildly sensitive to cyclic modulation but strongly depends on both compression force $f$ and duration $k$. The findings could assist the interpretation of \dmi{experiments using  fluorescence molecular tracers to track physical knots in polymers.} Furthermore, we identify optimal regions in the experimentally-controllable parameter space in which to obtain more/less topological and geometrical \dmi{entanglement}; this may \dmi{help designing} polymers with targeted topology.
\end{abstract}


\maketitle

Although it is generally well established that self-entanglements play vital roles in affecting polymer dynamics, their time evolution under strongly non equilibrium conditions, such us those achieved via electric fields, convergent flows or mechanical compression~\cite{Khorshid2014,Soh2019a,Tang2011,Suma2018,Amin2018}
are still poorly understood.  Under such conditions polymeric filaments cannot return to their equilibrium state until long after the perturbation has ended. The relevant relaxation time can be orders of magnitude longer than the characteristic Rouse timescales that are valid in equilibrium. For instance, $\mu$m-long DNA filaments can take seconds to relax after compression by electric pulses~\cite{Tang2011} or under confinement into narrow slits~\cite{balducci_et_al_prl_2007}.

\begin{figure*}[t!]
	\includegraphics[width=0.98\textwidth]{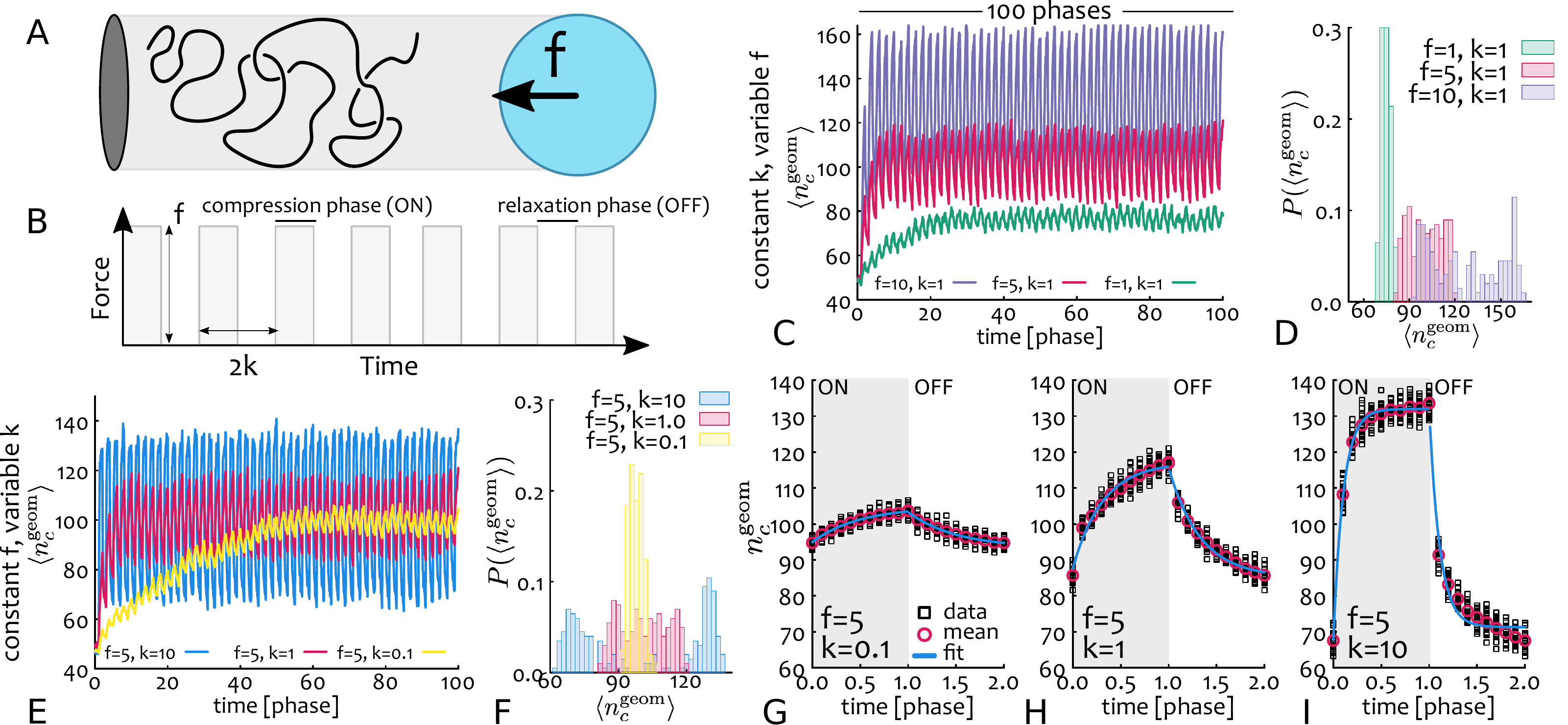}
		\vspace{-0.3cm}
	\caption{\textbf{A} Sketch of the system set up. \textbf{B} One compression-extension cycle consists of two equal phases, both of time $k$: the piston first exerts compressive force $f$ followed by a zero force recovery phase. \textbf{C,E} \dmi{The measure of geometrical complexity, $n_c^{\rm geom}$, is here averaged (as denoted by brackets) at equal times across simulation replicas at fixed $(f,k)$}. \textbf{D,F} Distribution of \dmi{$\langle n_c^{\rm geom} \rangle $} at steady state. Note that the mean value depends on $f$ but not $k$ whereas the variance depends on both. \textbf{G-I} \dmi{$\langle n_c^{\rm geom} \rangle$} in the ON and OFF phases for $f=5 k_BT/\sigma$ and $k=0.1, 1, 10$ $10^4 \tau_{LJ}$. \dmi{The average of $n_c^{\rm geom}$ is here additionally taken over the period. Note that time is given in units of cycles, which varies across panels.}  \dmi{$\langle n_c^{\rm geom} \rangle$} is fitted \dmi{with} the expression $d_0(1-\exp{\left( -t/\tau \right)})+d_1$ for the ON phase and $d'_0\exp{\left(-t/\tau\right)}+d'_1$ for the OFF phase with $\chi^2 \simeq 1$ for all fits in panels. In the legends, $k$ is in units of $10^6$ \dmi{integration} steps = $10^4$ $\tau_{LJ}$ and $f$ is in units of $k_B T/\sigma$ throughout this work.}
		\vspace{-0.5cm}
	\label{fig:methods}
\end{figure*}

Such unusually-slow relaxations have been associated with the emergence of complex forms of self-entanglement that hinder conformational rearrangements of molecules and trap them in long-lived states~\cite{Soh2019a,Michieletto2014self}. This view is supported by the observation of localised bright spots in fluorescence molecular traces, which can be interpreted as physical knots, that is knots in open chains~\cite{Soh2019a,Bao2003,mansfield1994there,Micheletti2011}.

Several central questions remain unanswered. How exactly does entanglement build up or decrease during non-equilibrium dynamics? What is its inherent degree of variability and spatial inhomogeneity? Is it feasible to control the out-of-equilibrium driving so to produce polymers with a target level of entanglement and heterogeneity?

In this computational study, we address these questions by considering a semiflexible polymer and confining it in a cylindrical channel where it undergoes compression and expansion phases (see Fig.~\ref{fig:methods}A). The compression is driven by a piston, a setup inspired by the ``nanodozer'' arrangement of refs.~\cite{Khorshid2014,Pelletier2012}.

By using various topological measures adapted to the case of linear  chains~\cite{Tubiana2011}, we systematically profile the nature and complexity of self-entanglements that arises in such system. We show that entanglement created by cyclic compression manifests in two different forms entanglement, geometrical and topological, that we can distinguish with suitable order parameters.
\dmi{We measure the complexity of topological entanglement in a linear chain, denoted by $n_c^{\rm topo}$, via the number of crossings in the simplest (minimal) diagrammatic representation of the knot trapped in the chain once its termini are suitably bridged~\cite{Tubiana2011prl,Tubiana2018}. The complexity of geometric entanglement, denoted by $n_c^{\rm geom}$, is instead defined in terms of the self-crossings in planar projections of the chain (without closure) and is computed as the number of self-crossings averaged over all projections~\cite{Freedman_et_al_AnnMath_1994}.}

\dmi{We discover that these two forms of entanglement evolve very differently: while the geometric one responds to compression/relaxation cycles similarly to driven dissipative systems, topological entanglement, reflecting the inherent complexity of physical knots, is only mildly coupled to
the instantaneous driving or to the geometrical entanglements and is long lived.} In fact, we show that significant physical knotting sets in only at suitable choices of the compressive force and period.
The system can therefore be steered towards the formation of more geometrical or topological entanglements  via the amplitude and period of the compressive force. Our findings open new perspectives for characterizing entanglement in fluctuating filaments and how to direct it by with extrinsic means.

\paragraph{The model --}

We consider a general model of a semi-flexible polymer confined in a cylinder stopped at one end by a hard planar wall. The polymer is periodically pushed against the wall by a spherical piston that occludes the cylinder cross-section, see Fig.~\ref{fig:methods}A. The chain is made of $N=1000$ beads of size $\sigma$ and has a nominal persistence length of $l_p=10\sigma$. The cylinder has radius $r=40 \sigma$ this being suitable to detect  knotting~\cite{Micheletti2014,uehara2019bimodality}. The system is evolved via Langevin dynamics simulations using standard values for the mass and friction coefficient~\cite{Kremer1990}(see SM).

In this initial study we only consider excluded volume interactions of the polymer with itself and with the piston (WCA repulsion) and the walls of the channel (quadratic repulsion), neglecting hydrodynamic effects. Including these would be computationally expensive and they are not expected to alter the ensuing entanglement properties~\cite{Marenda_et_al_soft_matt_2017,Weiss_et_al_macromol_2019}.

We focus on the case where the piston is moved by a periodic driving force parallel to the channel axis. The amplitude of the force oscillates between the values 0 and $f$ following a square wave with period $2k$ with each compression/extension phase having duration $k$ (which in this work we express in units of $10^6$ integration steps or $10^4$ $\tau_{LJ}$, see SI). We typically start from an equilibrated polymer conformation and drive the system for at least 100 compression/extension cycles in order to attain a steady state (see Fig.~\ref{fig:methods}C,E).

\paragraph{Cyclic evolution of geometrical entanglement --}
We take advantage of the structural details available in our simulations to profile entanglements through observables that complement the metric elongation measures, such as the gyration radius (see SM), that are accessible experimentally~\cite{Soh2019a,Tang2011}. We first measured the \dmi{geometric entanglement for each sampled conformation by averaging $n_c^{\rm geom}$ at equal times over different realizations of the noise (replicas). We shall denote these averages as $\langle n_c^{\rm geom} \rangle$.}

While $f$ and $k$ both affect significantly the distribution of \dmi{$\langle n_c^{\rm geom} \rangle$}  depends on the driving force $f$ but not on the cycle-time $k$ (see Fig.~\ref{fig:methods}D,F)  Instead, varying $k$ affects solely the width of the \dmi{$\langle n_c^{\rm geom} \rangle$} distribution (see Fig.~\ref{fig:methods}D). More specifically, we find that for fixed force $f$, the distribution of \dmi{$\langle n_c^{\rm geom} \rangle$} changes from unimodal to bimodal for long enough cycle times $k$ (Fig.~\ref{fig:methods}F). This effect,  that we shall rationalize below, suggests the presence of characteristic time-scales in the system's dynamics during which a driving force of certain amplitude must act in order to form entanglements.

That the mean average crossing number is independent of the cycle time $k$ can also be probed by averaging the \dmi{$\langle n_c^{\rm geom} \rangle$} over the compression/relaxation cycles at steady state (we consider the last 30 cycles). These data, shown in  Fig.~\ref{fig:methods}G-I,  appear to saturate for large choices of $k$, while their average over a period (compression and relaxation phases) remains constant. Note that the curves are reminiscent of classic dissipative systems, such as RC circuits or Kelvin-Voigt dashpot models. However, while in these classic systems the relaxation timescale is intrinsic, i.e.~uncoupled from the external drive here the relaxation timescale does depends on the drive, and particularly on $k$. This suggests the existence of a complex and possibly multiscale relaxation kinetics underlying the evolution of geometrical entanglement. We aim to explore this interesting aspect in future work.

The results in Fig.~\ref{fig:methods} indicate that two tunable parameters, $f$ and $k$, allow for separate control of the mean and variance of the geometric entanglement at steady state (see SM for further details).

\begin{figure}[t!]
	\includegraphics[width=0.45\textwidth]{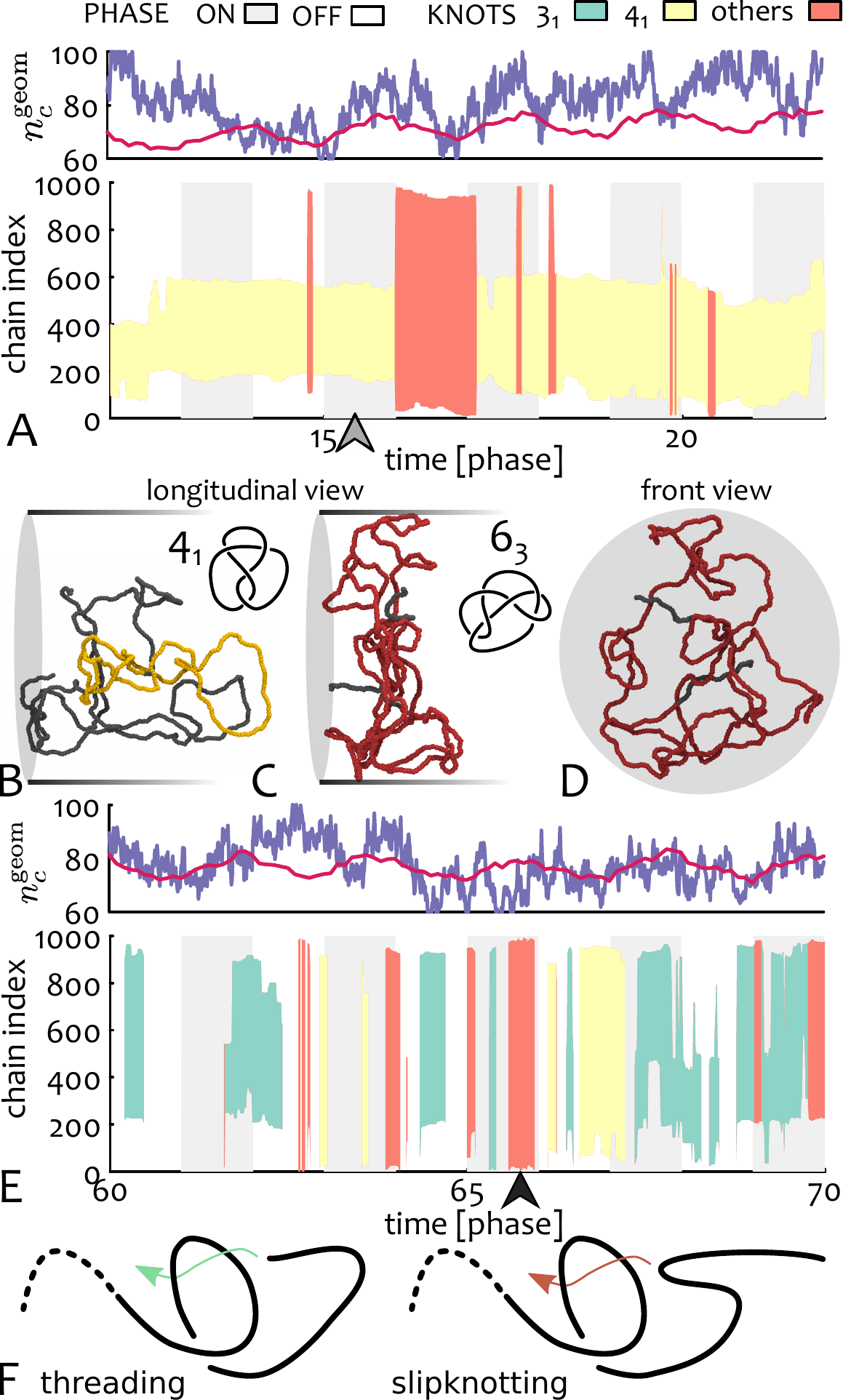}
	\vspace*{-0.2cm}
	\caption{\textbf{A,E} Kymographs for two time windows within the simulations with $f=1$ and $k=1$ showing the formation and evolution of knots. Knots are highly  dynamic and appear/disappear, even within a single compression/extension phase. Long-lived knots are less frequent. Note that most of the knots are shallow and spanning a large portion of the polymer (non-localised). The boundaries and topology of the knots are determined as median over a time-window of 11 time points equispaced by $\Delta t=10^3 \tau_{LJ}$ . The traces of \dmi{$n_c^{\rm geom}$} for this replica (blue) and the average over replicas (red) are shown at the top of the kymographs. \textbf{B,C,D} Three snapshots of two time points indicated with the grey (B) and black (C,D) arrow wedges. \textbf{F} Cartoons of knotting by threading (left) and slipknotting (right).}
	\vspace{-0.4cm}
	\label{fig:kymos}
\end{figure}

\begin{figure}[t!]
	\includegraphics[width=0.5\textwidth]{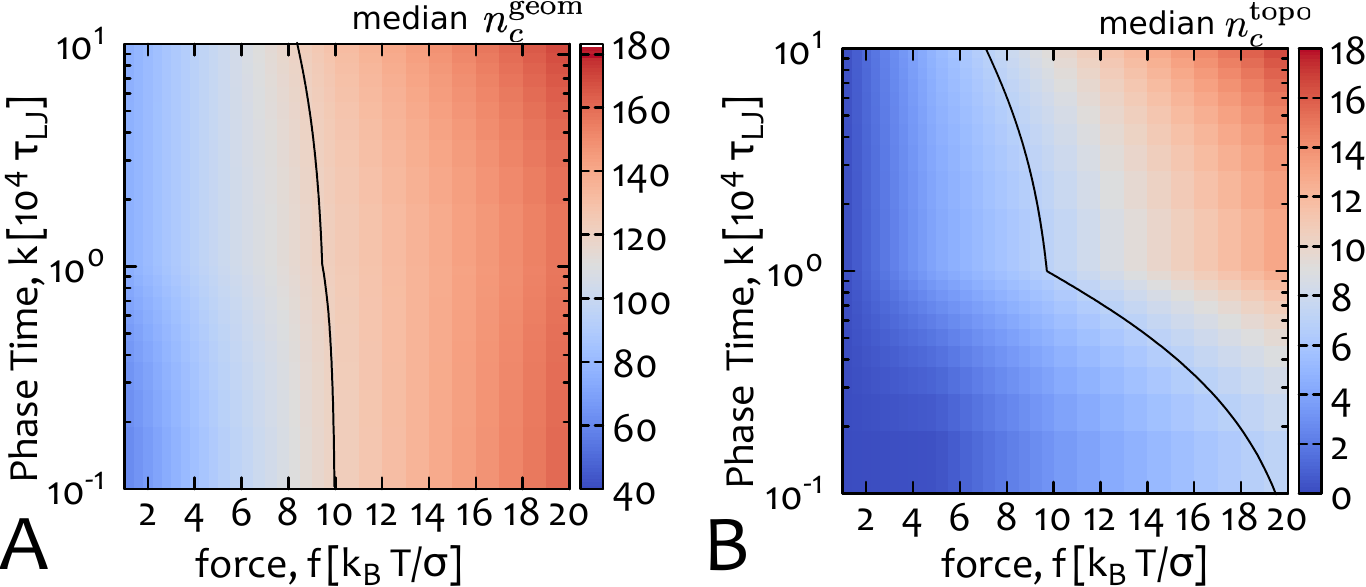}
	\vspace{-0.4cm}
	\caption{\dmi{Panels \textbf{A} and \textbf{B} show respectively the median $n_c^{\rm geom}$  and median $n_c^{\rm topo}$ observed} at steady state for various combinations of $f$ and $k$. These plots were obtained as linear interpolation of values obtained for $f=1,5,10,20$, $k=0.01, 0.1. 1, 10$. The black lines are the contours for median equal 120  (case A) and 6.7 (case B).}
		\vspace{-0.5cm}
	\label{fig:ncr_medians}
\end{figure}

\begin{figure*}[t!]
	\includegraphics[width=0.95\textwidth]{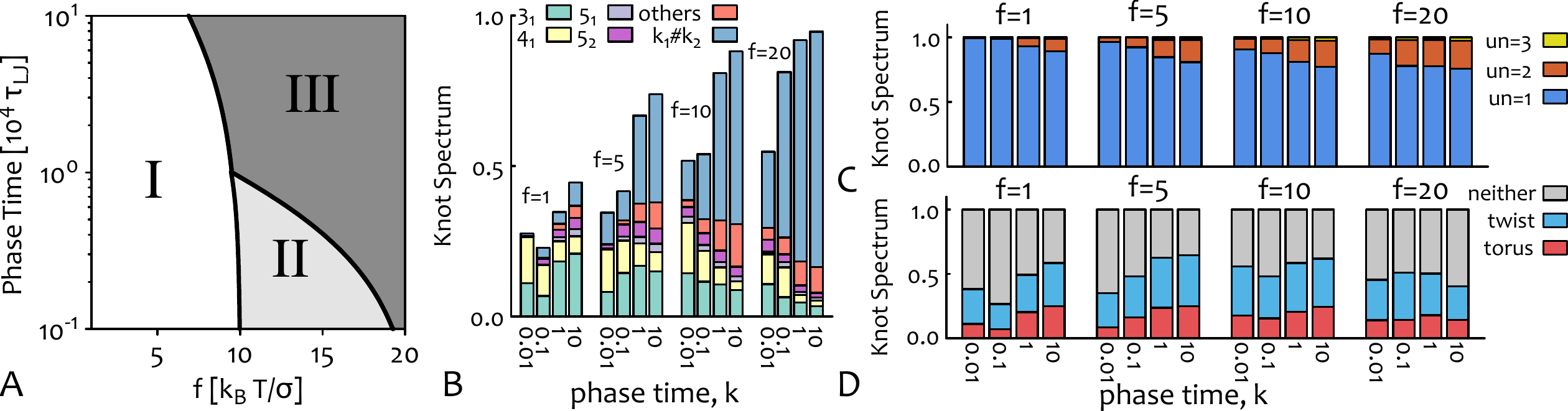}
	\vspace{-0.2cm}
	\caption{\textbf{A} Sketch of the state diagram identifying regions of the $(f,k)$ parameter space yielding different combinations of geometric and topological entanglement at steady state, see Fig~\ref{fig:ncr_medians}. Region I: low geometrical and topological \dmi{entanglement}. Region II: High geometrical and low topological \dmi{entanglement}. Region III: high geometrical and topological \dmi{entanglement}.
\textbf{B} Knot spectrum for different combinations $(f,k)$. \dmi{The group ``others'' contains non-trivial knots that, after topology-preserving simplifications, could not be assigned to the labelled groups.} \textbf{C} Unknotting number and \textbf{D} abundance of twist and torus knots.  In \textbf{C} and \textbf{D} we restricted (and normalised) the counting to prime knots with crossing number less or equal than 9 for simplicity. In \textbf{D}, the trefoil knot contributes to both twist and torus families. In this plot (and throughout the work) $k$ is given in units of $10^6$ \dmi{integration} steps = $10^4$ $\tau_{LJ}$ and $f$ is in units of $k_B T/\sigma$.}
	\vspace{-0.5cm}
	\label{fig:heatmap_spectrum}
\end{figure*}

\paragraph{Evolution of physical knots --}
To characterize the entanglement beyond its geometrical manifestation, as captured by \dmi{$\langle n_c^{\rm geom} \rangle$}, we tracked the formation of physical knots using the kymoknot software package~\cite{Tubiana2018}. This uses a minimally-invasive procedure to close an open chain~\cite{Tubiana2011} \dmi{by joining the termini either directly or through the convex hull of the chain, depending on which of the two provides the shortest route (see SI for more detail). The procedure allows us to assign a topological state (knot) to an open chain, locate the knotted portion along the chain contour and measure its complexity, e.g. via its minimal crossing number. }

Typical time evolutions of the knotted portion are presented in the kymographs of Fig.~\ref{fig:kymos}. Different colors are used to distinguish the various knot types that form, evolve and untie during the compression-expansion cycles. The colored region indicates the physical location of the knot on the chain contour. The temporal traces of the associated \dmi{$\langle n_c^{\rm geom} \rangle$} are also shown with the ``ON'' phase of the periodic compression (force) are marked by grey shading.

Physical knots clearly follow a rather different evolution from \dmi{$\langle n_c^{\rm geom} \rangle$}. After the initial transient (Fig.~\ref{fig:kymos}A) the \dmi{$\langle n_c^{\rm geom} \rangle$} varies continuously and in register with compression and expansion phases (Fig.~\ref{fig:kymos}E), retracing stochastically the loading and unloading curves of Fig.~\ref{fig:methods}C,D. Instead, over the same timescales, both the type and location of physical knots vary discontinuously and can persist across cycles. In fact, physical knots appear, disappear or change in complexity during both the compression and expansion phases.

The kymograph of Fig.~\ref{fig:kymos}A gives a striking illustration of the persistence of physical knots, specifically a $4_1$ knot that forms during the initial compression cycles and onto which additional transient entanglement is repeatedly introduced from either or both ends. We surmise that these types of long-lived knots are those observed in experiments using fluorescent microscopy, particularly the ones whose essential crossings coalesce in a tangle tight enough to create a bright fluorescence spot. Note that the tightness of the tangle does not necessarily correlate with the localization of the overall knotted region~\cite{Suma2017,Orlandini2019}. The trajectories of Fig.~\ref{fig:kymos} clarify that the whole knotted region is, in fact, delocalised, as it spans a significant portion of the chain.

Inspection of the boundaries of the knotted region reveals that the most common knot tying/untying mechanisms occurs at one, or both, chain ends, hence involving either the threading/unthreading of loops by the chain ends, or ``backfolding'' (Fig.~\ref{fig:kymos}F). In contrast, we rarely observe ``slip-knotting'' which occurs when a loop is threaded by a backfolded end. These events would manifest as a sudden appearance or disappearance of entanglement inside the chain, i.e. away from the chain ends~\cite{king2007identification,vskrbic2012role,sulkowska2013,Micheletti2014}. The overall behaviour is consistent with previously reported results for the spontaneous knotting of free chains in equilibrium~\cite{Tubiana2013a}.

\paragraph{Controlling the two forms of entanglement --}
\dmi{To quantify the external control on the geometrical and topological entanglement we profiled the dependence of the median values of $n_c^{\rm geom}$ and $n_c^{\rm topo}$ on $f$ and $k$ (see SM for details).  $n_c^{\rm topo}$ is the so-called crossing number and is a conventional measure of knot complexity. It corresponds to the number of essential crossings (a diagrammatic crossing is essential if its flipping would change the knot type) or, equivalently, to the number of crossings in the minimal diagrammatic representation of a given knot.
To establish it, we simplified the crossing pattern of a (closed) chain configuration first geometrically, using topology-preserving Monte Carlo moves~\cite{Koniaris1991,Taylor:2000:Nature:10972297}, and then symbolically via reductions and factorizations of the resulting Dowker code~\cite{knotscape}.} The latter was finally compared against a lookup table of knots with up to 16 crossings~\cite{knotscape}. The procedure allowed for establishing the precise knotted state and hence \dmi{$n_c^{\rm topo}$} of most of the sampled conformations, and used the lowest number of crossings after simplifications for the rest. These highly complex configurations were a minority of those sampled at any $(f,k)$ and therefore do not significantly bias the median \dmi{$n_c^{\rm topo}$} at any $(f,k)$ combination. The results are shown in Fig.~\ref{fig:ncr_medians}.

Here we see that the median \dmi{$n_c^{\rm topo}$} depends on both $f$ and fixed $k$. Instead, and consistently with the results of Fig.~\ref{fig:methods}C-D, the median \dmi{$ n_c^{\rm geom}$} depends on $f$ but only weakly, if at all, on $k$.
This is summarised in Fig.~\ref{fig:heatmap_spectrum}A. The state diagram shows a tripartite division of parameter space in which different combinations of geometrical and topological \dmi{entanglement} can be achieved by tuning $f$ and $k$.

\paragraph{Out-of-equilibrium knot spectrum --}
Our topological profiling strategy allows for pinpointing which types of physical knots are preferentially formed at the different values of $f$ and $k$. It  also provides a useful term of reference for future experiments aiming at studying polymers with a given topological entanglement. In fact, at small forces, an increase of the cycle time $k$ leads to an an increase of the simplest types of knots, particularly trefoil knots. However, at large forces, the population of trefoil knots decreases with $k$. The abundance of composite knots is also enhanced at larger forces, even for small periods while no composites are observed at small force (Fig.~\ref{fig:heatmap_spectrum}B). Both aspects are reminescent of knot populations in equilibrated chains or rings, which vary non-monotonically with increasing length~\cite{deguchi1993topology,katritch2000tightness,Millett2005,Tubiana2013a,Micheletti2012a}, with the force playing an analogous role as ring length; at the same time, the cycle time $k$ is a significant control parameter that has no analogue in equilibrated systems.

It is particularly interesting that most of the knots generated at $f=1$ and $k=\{0.01,0.1\}$ have unknotting number equal to 1, see Fig.~\ref{fig:heatmap_spectrum}C, meaning that they can be untied by a single suitable strand crossing. At the same time, the balance of twist and torus knots, shown in Fig.~\ref{fig:heatmap_spectrum}D, shows a slight dominance of torus knots, contrary to what has been observed in unconstrained equilibrated rings.

\paragraph{Conclusions -- }

Innovative single-molecule techniques have made it possible to explore the out-of-equilibrium behaviour of polymers by monitoring the fluorescence signals, and particularly the appearance of bright fluorescence spots, a likely signature of localised entanglement~\cite{Khorshid2014,Amin2018}. At the same time, key aspects of the out-of-equilibrium entanglement kinetics remains beyond the reach of such techniques and hence are still limitedly understood. For instance, genuine physical knots and other type of entanglement can both generate fluorescence spots~\cite{Jain2017a}. Additionally, the incidence of delocalised physical knots is not reliably known, as they would be challenging to detect from fluorescence signals. Accordingly, we still have an incomplete knowledge of the entanglement that inevitably forms in polymers driven out of equilibrium and how its evolution affects a polymer's behaviour.

To clarify these questions  we modelled a so-called ``nanodozer'' assay where a semiflexible chain is subject to cyclic compressionsinside a cylinder. For this prototypic system we  precisely monitored the formation and evolution of both geometrical and topological  entanglement,
the latter identified through the algorithmic detection of physical knots, regardless of how ``tight'' they were~\cite{Tubiana2011}.

We discover that these two types of entanglements (geometrical and topological) are broadly independent and evolving with different dynamics. For instance, the former follows closely the compression driving (Fig.~\ref{fig:methods}) whereas the latter does not (Fig.~\ref{fig:kymos}). We also observe that the interplay between the driving force and its period gives rise to a rich phenomenology; for instance, we clearly identify three regions in the parameter space with different levels of geometrical and topological (median) entanglement and also heterogeneity (Fig.~\ref{fig:ncr_medians}-\ref{fig:heatmap_spectrum}).
Importantly, we find that in some regions of the parameter space, these two aspects are decoupled and can therefore be independently tuned to generate of polymer states with desired combinations of geometrical and topological entanglement.

Our discoveries ought to be useful both in interpreting experimental results on out-of-equilibrium polymer entanglement and in designing new techniques to generate linear polymers with specific entanglement properties. Finally, the fact that many of the physical knots observed in our  simulations are delocalised (i.e. not tight, see Fig.~\ref{fig:kymos}), raises questions about whether these states can be faithfully detected in today's experimental assays or whether new experimental approaches are needed.

\paragraph{Acknowledgements. -- } The authors would like to acknowledge networking support by the COST Action CA17139. MST Acknowledges the generous support of the JSPS, via a long term fellowship, and the peerless hospitality of Prof Yamamoto at Kyoto University.
DM is supported by the Leverhulme Trust through an Early Career Fellowship (ECF-2019-088).\\

\bibliographystyle{apsrev4-1}
\bibliography{library,aux_lib}

\begin{thebibliography}{34}%
\makeatletter
\providecommand \@ifxundefined [1]{%
 \@ifx{#1\undefined}
}%
\providecommand \@ifnum [1]{%
 \ifnum #1\expandafter \@firstoftwo
 \else \expandafter \@secondoftwo
 \fi
}%
\providecommand \@ifx [1]{%
 \ifx #1\expandafter \@firstoftwo
 \else \expandafter \@secondoftwo
 \fi
}%
\providecommand \natexlab [1]{#1}%
\providecommand \enquote  [1]{``#1''}%
\providecommand \bibnamefont  [1]{#1}%
\providecommand \bibfnamefont [1]{#1}%
\providecommand \citenamefont [1]{#1}%
\providecommand \href@noop [0]{\@secondoftwo}%
\providecommand \href [0]{\begingroup \@sanitize@url \@href}%
\providecommand \@href[1]{\@@startlink{#1}\@@href}%
\providecommand \@@href[1]{\endgroup#1\@@endlink}%
\providecommand \@sanitize@url [0]{\catcode `\\12\catcode `\$12\catcode
  `\&12\catcode `\#12\catcode `\^12\catcode `\_12\catcode `\%12\relax}%
\providecommand \@@startlink[1]{}%
\providecommand \@@endlink[0]{}%
\providecommand \url  [0]{\begingroup\@sanitize@url \@url }%
\providecommand \@url [1]{\endgroup\@href {#1}{\urlprefix }}%
\providecommand \urlprefix  [0]{URL }%
\providecommand \Eprint [0]{\href }%
\providecommand \doibase [0]{http://dx.doi.org/}%
\providecommand \selectlanguage [0]{\@gobble}%
\providecommand \bibinfo  [0]{\@secondoftwo}%
\providecommand \bibfield  [0]{\@secondoftwo}%
\providecommand \translation [1]{[#1]}%
\providecommand \BibitemOpen [0]{}%
\providecommand \bibitemStop [0]{}%
\providecommand \bibitemNoStop [0]{.\EOS\space}%
\providecommand \EOS [0]{\spacefactor3000\relax}%
\providecommand \BibitemShut  [1]{\csname bibitem#1\endcsname}%
\let\auto@bib@innerbib\@empty
\bibitem [{\citenamefont {Khorshid}\ \emph {et~al.}(2014)\citenamefont
  {Khorshid}, \citenamefont {Zimny}, \citenamefont {Roche}, \citenamefont
  {Massarelli}, \citenamefont {Sakaue},\ and\ \citenamefont
  {Reisner}}]{Khorshid2014}%
  \BibitemOpen
  \bibfield  {author} {\bibinfo {author} {\bibfnamefont {A.}~\bibnamefont
  {Khorshid}}, \bibinfo {author} {\bibfnamefont {P.}~\bibnamefont {Zimny}},
  \bibinfo {author} {\bibfnamefont {D.~T.-l.}\ \bibnamefont {Roche}}, \bibinfo
  {author} {\bibfnamefont {G.}~\bibnamefont {Massarelli}}, \bibinfo {author}
  {\bibfnamefont {T.}~\bibnamefont {Sakaue}}, \ and\ \bibinfo {author}
  {\bibfnamefont {W.}~\bibnamefont {Reisner}},\ }\href@noop {} {\bibfield
  {journal} {\bibinfo  {journal} {Phys. Rev. Lett.}\ }\textbf {\bibinfo
  {volume} {113}},\ \bibinfo {pages} {268104} (\bibinfo {year}
  {2014})}\BibitemShut {NoStop}%
\bibitem [{\citenamefont {Soh}\ \emph {et~al.}(2019)\citenamefont {Soh},
  \citenamefont {Klotz}, \citenamefont {Robertson-Anderson},\ and\
  \citenamefont {Doyle}}]{Soh2019a}%
  \BibitemOpen
  \bibfield  {author} {\bibinfo {author} {\bibfnamefont {B.~W.}\ \bibnamefont
  {Soh}}, \bibinfo {author} {\bibfnamefont {A.~R.}\ \bibnamefont {Klotz}},
  \bibinfo {author} {\bibfnamefont {R.~M.}\ \bibnamefont {Robertson-Anderson}},
  \ and\ \bibinfo {author} {\bibfnamefont {P.~S.}\ \bibnamefont {Doyle}},\
  }\href@noop {} {\bibfield  {journal} {\bibinfo  {journal} {Physical Review
  Letters}\ }\textbf {\bibinfo {volume} {123}},\ \bibinfo {pages} {1} (\bibinfo
  {year} {2019})}\BibitemShut {NoStop}%
\bibitem [{\citenamefont {Tang}\ \emph {et~al.}(2011)\citenamefont {Tang},
  \citenamefont {Du},\ and\ \citenamefont {Doyle}}]{Tang2011}%
  \BibitemOpen
  \bibfield  {author} {\bibinfo {author} {\bibfnamefont {J.}~\bibnamefont
  {Tang}}, \bibinfo {author} {\bibfnamefont {N.}~\bibnamefont {Du}}, \ and\
  \bibinfo {author} {\bibfnamefont {P.~S.}\ \bibnamefont {Doyle}},\ }\href@noop
  {} {\bibfield  {journal} {\bibinfo  {journal} {Proc. Natl. Acad. Sci. USA}\
  }\textbf {\bibinfo {volume} {108}},\ \bibinfo {pages} {16153} (\bibinfo
  {year} {2011})}\BibitemShut {NoStop}%
\bibitem [{\citenamefont {Suma}\ \emph {et~al.}(2018)\citenamefont {Suma},
  \citenamefont {{Di Stefano}},\ and\ \citenamefont {Micheletti}}]{Suma2018}%
  \BibitemOpen
  \bibfield  {author} {\bibinfo {author} {\bibfnamefont {A.}~\bibnamefont
  {Suma}}, \bibinfo {author} {\bibfnamefont {M.}~\bibnamefont {{Di Stefano}}},
  \ and\ \bibinfo {author} {\bibfnamefont {C.}~\bibnamefont {Micheletti}},\
  }\href@noop {} {\bibfield  {journal} {\bibinfo  {journal} {Macromolecules}\
  }\textbf {\bibinfo {volume} {51}},\ \bibinfo {pages} {4462} (\bibinfo {year}
  {2018})}\BibitemShut {NoStop}%
\bibitem [{\citenamefont {Amin}\ \emph {et~al.}(2018)\citenamefont {Amin},
  \citenamefont {Khorshid}, \citenamefont {Zeng}, \citenamefont {Zimny},\ and\
  \citenamefont {Reisner}}]{Amin2018}%
  \BibitemOpen
  \bibfield  {author} {\bibinfo {author} {\bibfnamefont {S.}~\bibnamefont
  {Amin}}, \bibinfo {author} {\bibfnamefont {A.}~\bibnamefont {Khorshid}},
  \bibinfo {author} {\bibfnamefont {L.}~\bibnamefont {Zeng}}, \bibinfo {author}
  {\bibfnamefont {P.}~\bibnamefont {Zimny}}, \ and\ \bibinfo {author}
  {\bibfnamefont {W.}~\bibnamefont {Reisner}},\ }\href
  {http://dx.@doi.org/10.1038/s41467-018-03901-w} {\bibfield  {journal}
  {\bibinfo  {journal} {Nature Communications}\ }\textbf {\bibinfo {volume}
  {9}} (\bibinfo {year} {2018})}\BibitemShut {NoStop}%
\bibitem [{\citenamefont {Balducci}\ \emph {et~al.}(2007)\citenamefont
  {Balducci}, \citenamefont {Hsieh},\ and\ \citenamefont
  {Doyle}}]{balducci_et_al_prl_2007}%
  \BibitemOpen
  \bibfield  {author} {\bibinfo {author} {\bibfnamefont {A.}~\bibnamefont
  {Balducci}}, \bibinfo {author} {\bibfnamefont {C.-C.}\ \bibnamefont {Hsieh}},
  \ and\ \bibinfo {author} {\bibfnamefont {P.~S.}\ \bibnamefont {Doyle}},\
  }\href {http://link.aps.org/abstract/PRL/v99/e238102} {\bibfield  {journal}
  {\bibinfo  {journal} {Physical Review Letters}\ }\textbf {\bibinfo {volume}
  {99}},\ \bibinfo {eid} {238102} (\bibinfo {year} {2007})}\BibitemShut
  {NoStop}%
\bibitem [{\citenamefont {Michieletto}\ \emph {et~al.}(2014)\citenamefont
  {Michieletto}, \citenamefont {Marenduzzo}, \citenamefont {Orlandini},
  \citenamefont {Alexander},\ and\ \citenamefont
  {Turner}}]{Michieletto2014self}%
  \BibitemOpen
  \bibfield  {author} {\bibinfo {author} {\bibfnamefont {D.}~\bibnamefont
  {Michieletto}}, \bibinfo {author} {\bibfnamefont {D.}~\bibnamefont
  {Marenduzzo}}, \bibinfo {author} {\bibfnamefont {E.}~\bibnamefont
  {Orlandini}}, \bibinfo {author} {\bibfnamefont {G.~P.}\ \bibnamefont
  {Alexander}}, \ and\ \bibinfo {author} {\bibfnamefont {M.~S.}\ \bibnamefont
  {Turner}},\ }\href {http://dx.@doi.org/10.1039/C4SM00619D} {\bibfield
  {journal} {\bibinfo  {journal} {Soft Matter}\ }\textbf {\bibinfo {volume}
  {10}},\ \bibinfo {pages} {5936} (\bibinfo {year} {2014})}\BibitemShut
  {NoStop}%
\bibitem [{\citenamefont {Bao}\ \emph {et~al.}(2003)\citenamefont {Bao},
  \citenamefont {Lee},\ and\ \citenamefont {Quake}}]{Bao2003}%
  \BibitemOpen
  \bibfield  {author} {\bibinfo {author} {\bibfnamefont {X.~R.}\ \bibnamefont
  {Bao}}, \bibinfo {author} {\bibfnamefont {H.~J.}\ \bibnamefont {Lee}}, \ and\
  \bibinfo {author} {\bibfnamefont {S.~R.}\ \bibnamefont {Quake}},\ }\href@noop
  {} {\bibfield  {journal} {\bibinfo  {journal} {Phys. Rev. Lett.}\ }\textbf
  {\bibinfo {volume} {91}},\ \bibinfo {pages} {265506} (\bibinfo {year}
  {2003})}\BibitemShut {NoStop}%
\bibitem [{\citenamefont {Mansfield}(1994)}]{mansfield1994there}%
  \BibitemOpen
  \bibfield  {author} {\bibinfo {author} {\bibfnamefont {M.~L.}\ \bibnamefont
  {Mansfield}},\ }\href@noop {} {\bibfield  {journal} {\bibinfo  {journal}
  {Nature structural biology}\ }\textbf {\bibinfo {volume} {1}},\ \bibinfo
  {pages} {213} (\bibinfo {year} {1994})}\BibitemShut {NoStop}%
\bibitem [{\citenamefont {Micheletti}\ \emph {et~al.}(2011)\citenamefont
  {Micheletti}, \citenamefont {Marenduzzo},\ and\ \citenamefont
  {Orlandini}}]{Micheletti2011}%
  \BibitemOpen
  \bibfield  {author} {\bibinfo {author} {\bibfnamefont {C.}~\bibnamefont
  {Micheletti}}, \bibinfo {author} {\bibfnamefont {D.}~\bibnamefont
  {Marenduzzo}}, \ and\ \bibinfo {author} {\bibfnamefont {E.}~\bibnamefont
  {Orlandini}},\ }\href
  {http://linkinghub.elsevier.com/retrieve/pii/S0370157311000640} {\bibfield
  {journal} {\bibinfo  {journal} {Phys. Rep.}\ }\textbf {\bibinfo {volume}
  {504}},\ \bibinfo {pages} {1} (\bibinfo {year} {2011})}\BibitemShut {NoStop}%
\bibitem [{\citenamefont {Pelletier}\ \emph {et~al.}(2012)\citenamefont
  {Pelletier}, \citenamefont {Halvorsen}, \citenamefont {Ha}, \citenamefont
  {Paparcone}, \citenamefont {Sandler}, \citenamefont {Woldringh},
  \citenamefont {Wong},\ and\ \citenamefont {Jun}}]{Pelletier2012}%
  \BibitemOpen
  \bibfield  {author} {\bibinfo {author} {\bibfnamefont {J.}~\bibnamefont
  {Pelletier}}, \bibinfo {author} {\bibfnamefont {K.}~\bibnamefont
  {Halvorsen}}, \bibinfo {author} {\bibfnamefont {B.~Y.}\ \bibnamefont {Ha}},
  \bibinfo {author} {\bibfnamefont {R.}~\bibnamefont {Paparcone}}, \bibinfo
  {author} {\bibfnamefont {S.~J.}\ \bibnamefont {Sandler}}, \bibinfo {author}
  {\bibfnamefont {C.~L.}\ \bibnamefont {Woldringh}}, \bibinfo {author}
  {\bibfnamefont {W.~P.}\ \bibnamefont {Wong}}, \ and\ \bibinfo {author}
  {\bibfnamefont {S.}~\bibnamefont {Jun}},\ }\href@noop {} {\bibfield
  {journal} {\bibinfo  {journal} {Proceedings of the National Academy of
  Sciences of the United States of America}\ }\textbf {\bibinfo {volume} {109}}
  (\bibinfo {year} {2012})}\BibitemShut {NoStop}%
\bibitem [{\citenamefont {Tubiana}\ \emph
  {et~al.}(2011{\natexlab{a}})\citenamefont {Tubiana}, \citenamefont
  {Orlandini},\ and\ \citenamefont {Micheletti}}]{Tubiana2011}%
  \BibitemOpen
  \bibfield  {author} {\bibinfo {author} {\bibfnamefont {L.}~\bibnamefont
  {Tubiana}}, \bibinfo {author} {\bibfnamefont {E.}~\bibnamefont {Orlandini}},
  \ and\ \bibinfo {author} {\bibfnamefont {C.}~\bibnamefont {Micheletti}},\
  }\href@noop {} {\bibfield  {journal} {\bibinfo  {journal} {Prog. Theor. Phys.
  Suppl.}\ }\textbf {\bibinfo {volume} {191}},\ \bibinfo {pages} {192}
  (\bibinfo {year} {2011}{\natexlab{a}})}\BibitemShut {NoStop}%
\bibitem [{\citenamefont {Tubiana}\ \emph
  {et~al.}(2011{\natexlab{b}})\citenamefont {Tubiana}, \citenamefont
  {Orlandini},\ and\ \citenamefont {Micheletti}}]{Tubiana2011prl}%
  \BibitemOpen
  \bibfield  {author} {\bibinfo {author} {\bibfnamefont {L.}~\bibnamefont
  {Tubiana}}, \bibinfo {author} {\bibfnamefont {E.}~\bibnamefont {Orlandini}},
  \ and\ \bibinfo {author} {\bibfnamefont {C.}~\bibnamefont {Micheletti}},\
  }\href@noop {} {\bibfield  {journal} {\bibinfo  {journal} {Phys. Rev. Lett.}\
  }\textbf {\bibinfo {volume} {107}},\ \bibinfo {pages} {1} (\bibinfo {year}
  {2011}{\natexlab{b}})}\BibitemShut {NoStop}%
\bibitem [{\citenamefont {Tubiana}\ \emph {et~al.}(2018)\citenamefont
  {Tubiana}, \citenamefont {Polles}, \citenamefont {Orlandini},\ and\
  \citenamefont {Micheletti}}]{Tubiana2018}%
  \BibitemOpen
  \bibfield  {author} {\bibinfo {author} {\bibfnamefont {L.}~\bibnamefont
  {Tubiana}}, \bibinfo {author} {\bibfnamefont {G.}~\bibnamefont {Polles}},
  \bibinfo {author} {\bibfnamefont {E.}~\bibnamefont {Orlandini}}, \ and\
  \bibinfo {author} {\bibfnamefont {C.}~\bibnamefont {Micheletti}},\
  }\href@noop {} {\bibfield  {journal} {\bibinfo  {journal} {European Physical
  Journal E}\ }\textbf {\bibinfo {volume} {41}},\ \bibinfo {pages} {72}
  (\bibinfo {year} {2018})}\BibitemShut {NoStop}%
\bibitem [{\citenamefont {Freedman}\ \emph {et~al.}()\citenamefont {Freedman},
  \citenamefont {He},\ and\ \citenamefont
  {Wang}}]{Freedman_et_al_AnnMath_1994}%
  \BibitemOpen
  \bibfield  {author} {\bibinfo {author} {\bibfnamefont {M.}~\bibnamefont
  {Freedman}}, \bibinfo {author} {\bibfnamefont {Z.-X.}\ \bibnamefont {He}}, \
  and\ \bibinfo {author} {\bibfnamefont {Z.}~\bibnamefont {Wang}},\ }\href@noop
  {} {\bibfield  {journal} {\bibinfo  {journal} {Annals of Mathematics}\
  }\textbf {\bibinfo {volume} {139}},\ \bibinfo {pages} {1}}\BibitemShut
  {NoStop}%
\bibitem [{\citenamefont {Micheletti}\ and\ \citenamefont
  {Orlandini}(2014)}]{Micheletti2014}%
  \BibitemOpen
  \bibfield  {author} {\bibinfo {author} {\bibfnamefont {C.}~\bibnamefont
  {Micheletti}}\ and\ \bibinfo {author} {\bibfnamefont {E.}~\bibnamefont
  {Orlandini}},\ }\href {http://pubs.acs.org/@doi/abs/10.1021/mz500402s}
  {\bibfield  {journal} {\bibinfo  {journal} {ACS Macro Lett.}\ }\textbf
  {\bibinfo {volume} {3}},\ \bibinfo {pages} {876} (\bibinfo {year}
  {2014})}\BibitemShut {NoStop}%
\bibitem [{\citenamefont {Uehara}\ \emph {et~al.}(2019)\citenamefont {Uehara},
  \citenamefont {Coronel}, \citenamefont {Micheletti},\ and\ \citenamefont
  {Deguchi}}]{uehara2019bimodality}%
  \BibitemOpen
  \bibfield  {author} {\bibinfo {author} {\bibfnamefont {E.}~\bibnamefont
  {Uehara}}, \bibinfo {author} {\bibfnamefont {L.}~\bibnamefont {Coronel}},
  \bibinfo {author} {\bibfnamefont {C.}~\bibnamefont {Micheletti}}, \ and\
  \bibinfo {author} {\bibfnamefont {T.}~\bibnamefont {Deguchi}},\ }\href@noop
  {} {\bibfield  {journal} {\bibinfo  {journal} {Reactive and Functional
  Polymers}\ }\textbf {\bibinfo {volume} {134}},\ \bibinfo {pages} {141}
  (\bibinfo {year} {2019})}\BibitemShut {NoStop}%
\bibitem [{\citenamefont {Kremer}\ and\ \citenamefont
  {Grest}(1990)}]{Kremer1990}%
  \BibitemOpen
  \bibfield  {author} {\bibinfo {author} {\bibfnamefont {K.}~\bibnamefont
  {Kremer}}\ and\ \bibinfo {author} {\bibfnamefont {G.~S.}\ \bibnamefont
  {Grest}},\ }\href
  {http://link.aip.org/link/JCPSA6/v92/i8/p5057/s1{\&}Agg=@doi} {\bibfield
  {journal} {\bibinfo  {journal} {J. Chem. Phys.}\ }\textbf {\bibinfo {volume}
  {92}},\ \bibinfo {pages} {5057} (\bibinfo {year} {1990})}\BibitemShut
  {NoStop}%
\bibitem [{\citenamefont {Marenda}\ \emph {et~al.}(2017)\citenamefont
  {Marenda}, \citenamefont {Orlandini},\ and\ \citenamefont
  {Micheletti}}]{Marenda_et_al_soft_matt_2017}%
  \BibitemOpen
  \bibfield  {author} {\bibinfo {author} {\bibfnamefont {M.}~\bibnamefont
  {Marenda}}, \bibinfo {author} {\bibfnamefont {E.}~\bibnamefont {Orlandini}},
  \ and\ \bibinfo {author} {\bibfnamefont {C.}~\bibnamefont {Micheletti}},\
  }\href@noop {} {\bibfield  {journal} {\bibinfo  {journal} {Soft Matter}\
  }\textbf {\bibinfo {volume} {13}},\ \bibinfo {pages} {795} (\bibinfo {year}
  {2017})}\BibitemShut {NoStop}%
\bibitem [{\citenamefont {Weiss}\ \emph {et~al.}(2019)\citenamefont {Weiss},
  \citenamefont {Marenda}, \citenamefont {Micheletti},\ and\ \citenamefont
  {Likos}}]{Weiss_et_al_macromol_2019}%
  \BibitemOpen
  \bibfield  {author} {\bibinfo {author} {\bibfnamefont {L.}~\bibnamefont
  {Weiss}}, \bibinfo {author} {\bibfnamefont {M.}~\bibnamefont {Marenda}},
  \bibinfo {author} {\bibfnamefont {C.}~\bibnamefont {Micheletti}}, \ and\
  \bibinfo {author} {\bibfnamefont {C.}~\bibnamefont {Likos}},\ }\href@noop {}
  {\bibfield  {journal} {\bibinfo  {journal} {Macromolecules}\ }\textbf
  {\bibinfo {volume} {52}},\ \bibinfo {pages} {4111} (\bibinfo {year}
  {2019})}\BibitemShut {NoStop}%
\bibitem [{\citenamefont {Suma}\ and\ \citenamefont
  {Micheletti}(2017)}]{Suma2017}%
  \BibitemOpen
  \bibfield  {author} {\bibinfo {author} {\bibfnamefont {A.}~\bibnamefont
  {Suma}}\ and\ \bibinfo {author} {\bibfnamefont {C.}~\bibnamefont
  {Micheletti}},\ }\href@noop {} {\bibfield  {journal} {\bibinfo  {journal}
  {Proc. Nat. Acad. Sci. USA}\ ,\ \bibinfo {pages} {E2991}} (\bibinfo {year}
  {2017})}\BibitemShut {NoStop}%
\bibitem [{\citenamefont {Orlandini}\ \emph {et~al.}(2019)\citenamefont
  {Orlandini}, \citenamefont {Marenduzzo},\ and\ \citenamefont
  {Michieletto}}]{Orlandini2019}%
  \BibitemOpen
  \bibfield  {author} {\bibinfo {author} {\bibfnamefont {E.}~\bibnamefont
  {Orlandini}}, \bibinfo {author} {\bibfnamefont {D.}~\bibnamefont
  {Marenduzzo}}, \ and\ \bibinfo {author} {\bibfnamefont {D.}~\bibnamefont
  {Michieletto}},\ }\href
  {http://www.pnas.org/lookup/@doi/10.1073/pnas.1815394116} {\bibfield
  {journal} {\bibinfo  {journal} {Proceedings of the National Academy of
  Sciences}\ }\textbf {\bibinfo {volume} {116}},\ \bibinfo {pages} {8149}
  (\bibinfo {year} {2019})}\BibitemShut {NoStop}%
\bibitem [{\citenamefont {King}\ \emph {et~al.}(2007)\citenamefont {King},
  \citenamefont {Yeates},\ and\ \citenamefont
  {Yeates}}]{king2007identification}%
  \BibitemOpen
  \bibfield  {author} {\bibinfo {author} {\bibfnamefont {N.~P.}\ \bibnamefont
  {King}}, \bibinfo {author} {\bibfnamefont {E.~O.}\ \bibnamefont {Yeates}}, \
  and\ \bibinfo {author} {\bibfnamefont {T.~O.}\ \bibnamefont {Yeates}},\
  }\href@noop {} {\bibfield  {journal} {\bibinfo  {journal} {Journal of
  molecular biology}\ }\textbf {\bibinfo {volume} {373}},\ \bibinfo {pages}
  {153} (\bibinfo {year} {2007})}\BibitemShut {NoStop}%
\bibitem [{\citenamefont {{\v{S}}krbi{\'c}}\ \emph {et~al.}(2012)\citenamefont
  {{\v{S}}krbi{\'c}}, \citenamefont {Micheletti},\ and\ \citenamefont
  {Faccioli}}]{vskrbic2012role}%
  \BibitemOpen
  \bibfield  {author} {\bibinfo {author} {\bibfnamefont {T.}~\bibnamefont
  {{\v{S}}krbi{\'c}}}, \bibinfo {author} {\bibfnamefont {C.}~\bibnamefont
  {Micheletti}}, \ and\ \bibinfo {author} {\bibfnamefont {P.}~\bibnamefont
  {Faccioli}},\ }\href@noop {} {\bibfield  {journal} {\bibinfo  {journal} {PLoS
  computational biology}\ }\textbf {\bibinfo {volume} {8}} (\bibinfo {year}
  {2012})}\BibitemShut {NoStop}%
\bibitem [{\citenamefont {Su{\l}kowska}\ \emph {et~al.}(2013)\citenamefont
  {Su{\l}kowska}, \citenamefont {Noel}, \citenamefont {Ram{\'i}rez-Sarmiento},
  \citenamefont {Rawdon}, \citenamefont {Millett},\ and\ \citenamefont
  {Onuchic}}]{sulkowska2013}%
  \BibitemOpen
  \bibfield  {author} {\bibinfo {author} {\bibfnamefont {J.}~\bibnamefont
  {Su{\l}kowska}}, \bibinfo {author} {\bibfnamefont {J.}~\bibnamefont {Noel}},
  \bibinfo {author} {\bibfnamefont {C.}~\bibnamefont {Ram{\'i}rez-Sarmiento}},
  \bibinfo {author} {\bibfnamefont {E.}~\bibnamefont {Rawdon}}, \bibinfo
  {author} {\bibfnamefont {K.}~\bibnamefont {Millett}}, \ and\ \bibinfo
  {author} {\bibfnamefont {J.}~\bibnamefont {Onuchic}},\ }\href@noop {}
  {\bibfield  {journal} {\bibinfo  {journal} {Biochemical Society
  Transactions}\ }\textbf {\bibinfo {volume} {41}},\ \bibinfo {pages} {523}
  (\bibinfo {year} {2013})}\BibitemShut {NoStop}%
\bibitem [{\citenamefont {Tubiana}\ \emph {et~al.}(2013)\citenamefont
  {Tubiana}, \citenamefont {Rosa}, \citenamefont {Fragiacomo},\ and\
  \citenamefont {Micheletti}}]{Tubiana2013a}%
  \BibitemOpen
  \bibfield  {author} {\bibinfo {author} {\bibfnamefont {L.}~\bibnamefont
  {Tubiana}}, \bibinfo {author} {\bibfnamefont {A.}~\bibnamefont {Rosa}},
  \bibinfo {author} {\bibfnamefont {F.}~\bibnamefont {Fragiacomo}}, \ and\
  \bibinfo {author} {\bibfnamefont {C.}~\bibnamefont {Micheletti}},\ }\href
  {http://pubs.acs.org/@doi/abs/10.1021/ma4002963} {\bibfield  {journal}
  {\bibinfo  {journal} {Macromolecules}\ }\textbf {\bibinfo {volume} {46}},\
  \bibinfo {pages} {3669} (\bibinfo {year} {2013})}\BibitemShut {NoStop}%
\bibitem [{\citenamefont {Koniaris}\ and\ \citenamefont
  {Muthukumar}(1991)}]{Koniaris1991}%
  \BibitemOpen
  \bibfield  {author} {\bibinfo {author} {\bibfnamefont {K.}~\bibnamefont
  {Koniaris}}\ and\ \bibinfo {author} {\bibfnamefont {M.}~\bibnamefont
  {Muthukumar}},\ }\href
  {http://link.aip.org/link/JCPSA6/v95/i4/p2873/s1{\&}Agg=@doi} {\bibfield
  {journal} {\bibinfo  {journal} {The Journal of Chemical Physics}\ }\textbf
  {\bibinfo {volume} {95}},\ \bibinfo {pages} {2873} (\bibinfo {year}
  {1991})}\BibitemShut {NoStop}%
\bibitem [{\citenamefont {{Taylor, WR}}(2000)}]{Taylor:2000:Nature:10972297}%
  \BibitemOpen
  \bibfield  {author} {\bibinfo {author} {\bibnamefont {{Taylor, WR}}},\
  }\href@noop {} {\bibfield  {journal} {\bibinfo  {journal} {Nature}\ }\textbf
  {\bibinfo {volume} {406}},\ \bibinfo {pages} {916} (\bibinfo {year}
  {2000})}\BibitemShut {NoStop}%
\bibitem [{\citenamefont {Hoste}\ and\ \citenamefont
  {Thistlethwaite}()}]{knotscape}%
  \BibitemOpen
  \bibfield  {author} {\bibinfo {author} {\bibfnamefont {J.}~\bibnamefont
  {Hoste}}\ and\ \bibinfo {author} {\bibfnamefont {M.}~\bibnamefont
  {Thistlethwaite}},\ }\href@noop {} {\bibinfo  {journal}
  {http://www.math.utk.edu/~morwen/knotscape.html}\ }\BibitemShut {NoStop}%
\bibitem [{\citenamefont {Deguchi}\ and\ \citenamefont
  {Tsurusaki}(1993)}]{deguchi1993topology}%
  \BibitemOpen
\bibfield  {journal} {  }\bibfield  {author} {\bibinfo {author} {\bibfnamefont
  {T.}~\bibnamefont {Deguchi}}\ and\ \bibinfo {author} {\bibfnamefont
  {K.}~\bibnamefont {Tsurusaki}},\ }\href@noop {} {\bibfield  {journal}
  {\bibinfo  {journal} {Journal of the Physical Society of Japan}\ }\textbf
  {\bibinfo {volume} {62}},\ \bibinfo {pages} {1411} (\bibinfo {year}
  {1993})}\BibitemShut {NoStop}%
\bibitem [{\citenamefont {Katritch}\ \emph {et~al.}(2000)\citenamefont
  {Katritch}, \citenamefont {Olson}, \citenamefont {Vologodskii}, \citenamefont
  {Dubochet},\ and\ \citenamefont {Stasiak}}]{katritch2000tightness}%
  \BibitemOpen
  \bibfield  {author} {\bibinfo {author} {\bibfnamefont {V.}~\bibnamefont
  {Katritch}}, \bibinfo {author} {\bibfnamefont {W.~K.}\ \bibnamefont {Olson}},
  \bibinfo {author} {\bibfnamefont {A.}~\bibnamefont {Vologodskii}}, \bibinfo
  {author} {\bibfnamefont {J.}~\bibnamefont {Dubochet}}, \ and\ \bibinfo
  {author} {\bibfnamefont {A.}~\bibnamefont {Stasiak}},\ }\href@noop {}
  {\bibfield  {journal} {\bibinfo  {journal} {Physical Review E}\ }\textbf
  {\bibinfo {volume} {61}},\ \bibinfo {pages} {5545} (\bibinfo {year}
  {2000})}\BibitemShut {NoStop}%
\bibitem [{\citenamefont {Millett}\ \emph {et~al.}(2005)\citenamefont
  {Millett}, \citenamefont {Dobay},\ and\ \citenamefont
  {Stasiak}}]{Millett2005}%
  \BibitemOpen
  \bibfield  {author} {\bibinfo {author} {\bibfnamefont {K.}~\bibnamefont
  {Millett}}, \bibinfo {author} {\bibfnamefont {A.}~\bibnamefont {Dobay}}, \
  and\ \bibinfo {author} {\bibfnamefont {A.}~\bibnamefont {Stasiak}},\
  }\href@noop {} {\bibfield  {journal} {\bibinfo  {journal} {Macromolecules}\
  }\textbf {\bibinfo {volume} {38}},\ \bibinfo {pages} {601} (\bibinfo {year}
  {2005})}\BibitemShut {NoStop}%
\bibitem [{\citenamefont {Micheletti}\ and\ \citenamefont
  {Orlandini}(2012)}]{Micheletti2012a}%
  \BibitemOpen
  \bibfield  {author} {\bibinfo {author} {\bibfnamefont {C.}~\bibnamefont
  {Micheletti}}\ and\ \bibinfo {author} {\bibfnamefont {E.}~\bibnamefont
  {Orlandini}},\ }\href {http://xlink.rsc.org/?@doi=c2sm26401c} {\bibfield
  {journal} {\bibinfo  {journal} {Soft Matter}\ }\textbf {\bibinfo {volume}
  {8}},\ \bibinfo {pages} {10959} (\bibinfo {year} {2012})}\BibitemShut
  {NoStop}%
\bibitem [{\citenamefont {Jain}\ and\ \citenamefont
  {Dorfman}(2017)}]{Jain2017a}%
  \BibitemOpen
  \bibfield  {author} {\bibinfo {author} {\bibfnamefont {A.}~\bibnamefont
  {Jain}}\ and\ \bibinfo {author} {\bibfnamefont {K.~D.}\ \bibnamefont
  {Dorfman}},\ }\href {http://dx.@doi.org/10.1063/1.4979605} {\bibfield
  {journal} {\bibinfo  {journal} {Biomicrofluidics}\ }\textbf {\bibinfo
  {volume} {11}},\ \bibinfo {pages} {1} (\bibinfo {year} {2017})}\BibitemShut
  {NoStop}%
\end{thebibliography}%

\end{document}